\documentclass{JINST}
\newfont{\thiplo}{msbm10 scaled\magstep 2}
\newfont{\gothic}{eufb10 scaled\magstep 2}
\newfont{\unc}{eurb10} 
\newskip\humongous \humongous=0pt plus 1000pt minus 1000pt
\def\caja{\mathsurround=0pt}
\def\eqalign#1{\,\vcenter{\openup1\jot \caja
        \ialign{\strut \hfil$\displaystyle{##}$&$
        \displaystyle{{}##}$\hfil\crcr#1\crcr}}\,}
\newif\ifdtup

\def\eqright #1\cr{\noalign{\hfill$\displaystyle{{}#1}$}}
\def\eqleft #1\cr{\noalign{\noindent$\displaystyle{{}#1}$\hfill}}
\def\oldreffmt#1{\rlap{[#1]} \hbox to 2\parindent{}}

\def\figfmt#1{\rlap{Figure {#1}} \hbox to 1in{}}

\def\sectioneq{\def\theequation{\thesection.\arabic{equation}}{\let
\holdsection=\section\def\section{\setcounter{equation}{0}\holdsection}}}%

\newcounter{holdequation}

\def\begineq #1\endeq{$$ \refstepcounter{equation}\eqalign{#1}\eqno
	(\theequation) $$}
\def\contlimit{\,{\hbox{$\longrightarrow$}\kern-1.8em\lower1ex
\hbox{${\scriptstyle (a\rightarrow0)}$}}\,}
\def\centeron#1#2{{\setbox0=\hbox{#1}\setbox1=\hbox{#2}\ifdim
\wd1>\wd0\kern.5\wd1\kern-.5\wd0\fi
\copy0\kern-.5\wd0\kern-.5\wd1\copy1\ifdim\wd0>\wd1
\kern.5\wd0\kern-.5\wd1\fi}}
\def\centerover#1#2{\centeron{#1}{\setbox0=\hbox{#1}\setbox
1=\hbox{#2}\raise\ht0\hbox{\raise\dp1\hbox{\copy1}}}}
\def\centerunder#1#2{\centeron{#1}{\setbox0=\hbox{#1}\setbox
1=\hbox{#2}\lower\dp0\hbox{\lower\ht1\hbox{\copy1}}}}
\def\lsim{\;\centeron{\raise.35ex\hbox{$<$}}{\lower.65ex\hbox
{$\sim$}}\;}
\def\gsim{\;\centeron{\raise.35ex\hbox{$>$}}{\lower.65ex\hbox
{$\sim$}}\;}
\def\super#1{\ifmmode \hbox{\textsuper{#1}}\else\textsuper{#1}\fi}
\def\textsuper#1{\newcount\holdspacefactor\holdspacefactor=\spacefactor
$^{#1}$\spacefactor=\holdspacefactor}
\def\getcite#1,{\advance\citenumber by1
\def\getcitearg{#1}\def\lastarg{@}
\ifnum\citenumber=1
\ref{#1}\let\next=\getcite\else\ifx\getcitearg\lastarg\let\next=\relax
\else ,\ref{#1}\let\next=\getcite\fi\fi\next}
\def\pom{{\rm P\kern -0.53em\llap I\,}}
\def\spom{{\rm P\kern -0.36em\llap \small I\,}}
\def\sspom{{\rm P\kern -0.33em\llap \footnotesize I\,}}

\relax
\def\contlimit{\,{\hbox{$\longrightarrow$}\kern-1.8em\lower1ex
\hbox{${\scriptstyle (a\rightarrow0)}$}}\,}
\def\upon #1/#2 {{\textstyle{#1\over #2}}}
\relax

\sectioneq

\def\til#1{\centeron{\hbox{$#1$}}{\lower 2ex\hbox{$\char'176$}}}
\def\tild#1{\centeron{\hbox{$\,#1$}}{\lower 2.5ex\hbox{$\char'176$}}}
\def\sumtil{\centeron{\hbox{$\displaystyle\sum$}}{\lower
-1.5ex\hbox{$\widetilde{\phantom{xx}}$}}}

\rightline{\vbox{\halign{&#\hfil\cr
&\today\cr}}} 
\vspace{0.25in}

\title{Forward Physics with Rapidity Gaps at the LHC}

\author{M.G.Albrow$^a$\thanks{Corresponding
author.}, A.DeRoeck$^b$, V.A.Khoze$^c$, J.L\"{a}ms\"{a}$^{d,e}$, E.Norbeck$^f$,
Y.Onel$^f$, R.Orava$^e$, and M.G.Ryskin$^g$.\\
\llap{$^a$}Fermi National Accelerator Laboratory, Batavia, IL60510, U.S.A.\\
\llap{$^b$}CERN, CH-1211 Geneva, Switzerland.\\
  \llap{$^c$} IPPP, Dept. of Physics, Durham University, Durham, U.K. \\
  \llap{$^d$} Iowa State University, Ames, Iowa, U.S.A. \\
  \llap{$^e$} Dept. of Physics, Univ. of Helsinki, and Helsinki Inst. of Physics, Finland.\\
  \llap{$^f$} Univ. of Iowa, Iowa City, Iowa, U.S.A. \\
  \llap{$^g$} Petersburg Nuclear Physics Inst., Gatchina, St.Petersburg, Russia.

 E-mail: \email{albrow@fnal.gov,orava@mail.cern.ch}}

\abstract{A rapidity gap program with great potential can be realised at the Large Hadron Collider, LHC, by adding a few simple forward shower
counters (FSCs) along the beam line on both sides of the main central detectors, such as CMS. Measurements of single diffractive cross
sections down to the lowest masses can be made with an $\mathrm{eff}$icient level-1 trigger. Exceptionally, the detectors
also make feasible the study of Central Diffractive Excitation, and in particular the reaction $g+g \rightarrow
g + g$, in the color singlet channel, $\mathrm{ef}$fectively using the LHC as a gluon-gluon collider.}


\begin{document}

\section{Introduction}

\hspace{.5in}The principal goal of the program of forward physics at the Large Hadron Collider, LHC, is the measurement of the main
characteristics of diffractive interactions. These processes are very $\mathrm{signif}$icant in their own right, to better
understand QCD in the non-perturbative regime, and they form a large fraction of the total cross section. In
addition, they are valuable because of their intimate connection to the rapidity gap survival factor $\hat{S}^2$,
which determines the rate of suppression of the central exclusive processes caused by rescattering $\mathrm{ef}$fects
(including additional parton interactions). Such processes have a number of unique advantages for studying QCD
and new physics, and in particular for detailed probing of the Higgs boson properties~\cite{kmr1,mgafp420}.

For predictions of soft diffractive processes at the LHC we need a reliable theoretical model. There has not
been much improvement in the theoretical understanding of the high energy behaviour of the strong interaction
amplitude since the late 1960s~\cite{gribov}. For recent reviews, see Refs~\cite{kaid,rmks}. For instance, different asymptopic behaviours of such a fundamental quantity
as the total $pp$ interaction cross section, $\sigma_T$, are not excluded. Thus in the ``weak coupling'' regime,
$\sigma_T(s \rightarrow \infty) \rightarrow K$, where $K$ is a constant, while in the ``strong coupling'' regime
$\sigma_T \sim $ ln(s)$^\varepsilon$ with $0< \varepsilon < 2$. Note that for these two regimes different
behaviour of the cross section for diffractive dissociation is predicted. It is therefore important to
study different channels of diffractive dissociation at the LHC to make progress.

\begin{figure}[t]
  \vspace{9.0cm}
  \includegraphics{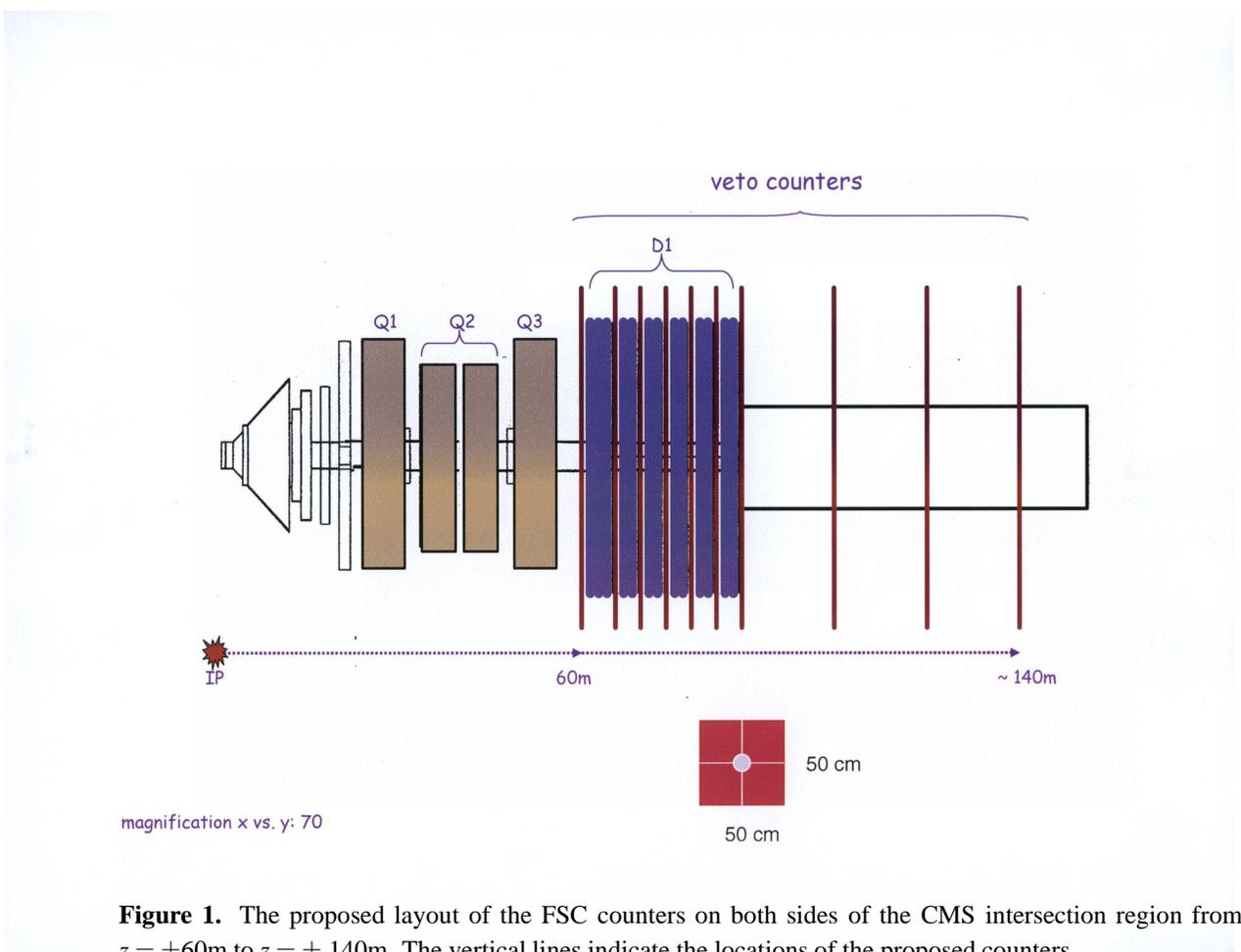}
    \vspace{-1.0cm}
  \caption{The proposed layout of the FSC counters on both sides of the CMS intersection region from $z = \pm$60m to $z =
  \pm$
  140m. The vertical lines indicate the locations of the proposed counters. }
%\label{fig:cdfpu}
\end{figure}

The experimental results available at present are fragmentary, and do not cover the whole kinematic range. 
To further constrain the parameters of the models of soft diffraction it is crucial to make accurate
measurements at LHC energies of the single diffractive dissociation cross section for low masses,
$\sigma_{SD}$(low $M$), and of central diffractive production, $\frac{d\sigma}{d\eta_1 d\eta_2}$, where $\eta_1$
and $\eta_2$ de$\mathrm{f}$ine the pseudorapidity ($\eta = -$ln tan$\frac{\theta}{2}$) range of the central system.

While the physics of diffractive dissociation at the LHC is important, the existing LHC detectors (ALICE, ATLAS, CMS, and LHCb) are
not well suited, as they lack the coverage necessary to measure forward rapidity gaps. The TOTEM experiment~\cite{totem,cmstotem}
 will measure
the total cross section and elastic scattering, detecting protons in Roman pots, and make some studies of high cross
section diffractive interactions. Downstream of the ATLAS experiment, the ALFA project consists of a set of Roman pots
with $\mathrm{f}$iber tracking, also for elastic scattering and total cross section measurements. In this paper, the concept of using forward
shower counters, FSC, to detect and trigger on rapidity gaps in diffractive events at the LHC is presented. The bene$\mathrm{f}$its of the FSC
system for the study of single and central diffractive reactions are shown. Extensive simulations of several reactions have been
performed to establish the $\mathrm{eff}$iciency of the FSC detector arrangement. The main application of these studies will be for low
luminosity running with a small probability of pile-up (more than one inelastic $pp$ collision per bunch crossing). This study parallels
that of Ref.~\cite{lamsa} and that presented in a proposal to the CMS Collaboration~\cite{albrowfsc}. In this paper we refer
speci$\mathrm{f}$ically to CMS, but the other three central detectors may similarly bene$\mathrm{f}$it from 
such very forward detectors.

\begin{figure}[t]
  \vspace{9.0cm}
  \includegraphics{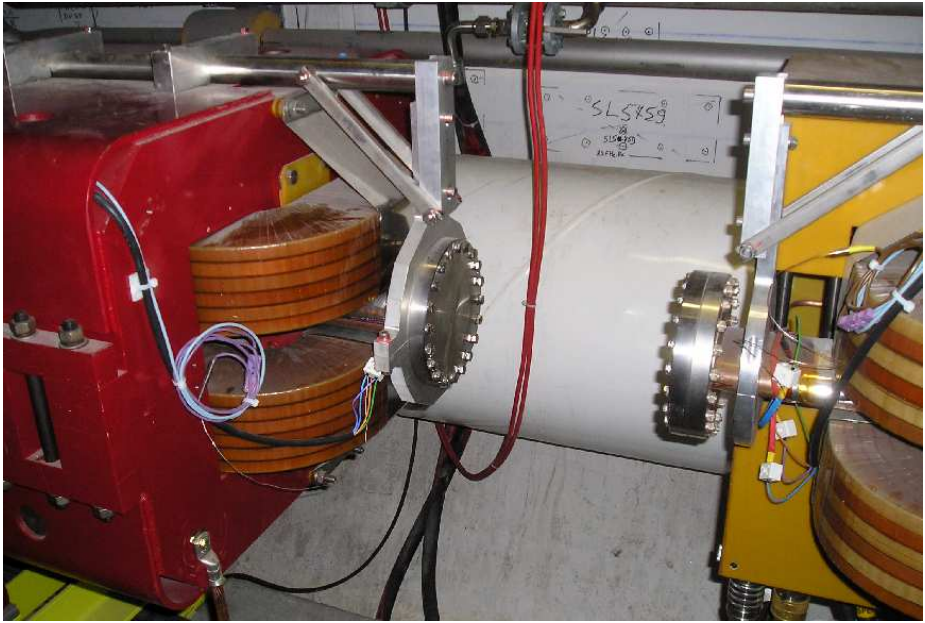}
  \caption{Region between the MBXW magnets where the FSC counters can be installed. (Photograph taken
  before the connecting pipe was installed; it shows accessible elliptical beam pipe regions.)}
%\label{fig:cdfpu}
\end{figure}

Scintillation counters can be added closely surrounding the beam pipes, with 60 m $<|z|<$ 85 m ($z$ is the coordinate along the beam
direction), and at further locations out
to $z = \pm$140 m on both sides of the interaction point (IP5), see Figure 1. At these locations the beam pipes are accessible, and not in a 
cryogenic region. An option is to supplement scintillation counters with gaseous
electron multipliers (GEMs), which would add spatial segmentation, and therefore shower position information. The FSC do not detect
particles directly from the collisions, but they detect showers from forward particles that interact in the beam pipe and
surrounding material. 
These detectors can be used to make measurements of rapidity gaps for events without pile-up, in the early
days of LHC running. The average number of inelastic collisions per bunch crossing, $\langle n \rangle$, for $\sigma_{inel}$ = 80
mb, $6\times 10^{11}$ protons per bunch, 156 bunches and $\beta^*$ = 2 m is $\langle n \rangle \sim$ 1.0, and then $e^{-\langle n
\rangle}$ = 0.368 is the fraction of bunch crossings with no inelastic collisions (and 36.8\% of diffractive, or any
other speci$\mathrm{f}$ic, collisions are not spoiled by
pile-up). These conditions give the highest rate of diffractive collisions without pile-up. When $\langle n \rangle$ = 2(3)
rapidity gap physics can still be done, but with lower $\mathrm{eff}$iciency, $\varepsilon$ = 13.5(5.0)\% respectively. As most of the
pile-up events will have forward particles giving showers in the FSC, they can be $\mathrm{eff}$ectively vetoed at the level-1 trigger,
increasing the $\mathrm{eff}$iciency of diffractive triggers. For single diffractive excitation one would require 
all the counters on one
side (in logical OR) to be consistent with noise. Showers will usually give a large pulse height, many times that of a minimum
ionizing particle (MIP), and are easily discriminated from noise. Off-line, multiple events in a bunch crossing also usually give more
than one primary vertex, as reconstructed from the excellent tracking capabilities of CMS. However 
low mass diffractive excitation can
have all particles at small polar angles, and in some cases will not have measured tracks, or they will not form a reconstructable
primary vertex.

For simplicity of discussion, consider level-1 triggers in con$\mathrm{f}$igurations of gaps (detectors consistent with noise) 
or particles (signals signi$\mathrm{f}$icantly above noise), in three rapidity regions: forward ``East" side, 
central, and forward ``West" side.
``Central" may e.g. mean some speci$\mathrm{f}$ied activity in $|\eta| <$ 3.
Call [000] all empty, [001] particles only forward(West), etc., thus eight possible con$\mathrm{f}$igurations. Uninteresting crossings
with no inelastic interaction are [000], and non-diffractive and pile-up events are [111]; these will dominate zero-bias
data and are not relevant for diffractive physics. It is interesting to trigger on [001]+[100], representing low-mass single
diffraction, [011]+[110] which is high-mass single diffraction, [101] which is double diffractive dissociation, and [010]
which is double pomeron exchange (central diffraction). These six triggers all involve large $(\gsim 5)$ rapidity gaps
and will contain no pile-up events, so their rate will decrease approximately exponentially once $\langle n \rangle \gsim 3$.

In addition to their value in triggering on and studying diffractive collisions without pile-up, the FSC counters can provide real-time monitoring of beam
conditions (beam halo) for both incoming and outgoing beams (which are both in the same pipe at these locations). The separation of
incoming and outgoing beams can be done by timing the scintillation counter signals at a few locations 
where their time separation is $\gsim$10 ns (the maximum being 12.5 ns). This is likely to be a useful beam diagnostic, especially in early LHC operation.

These detectors will allow the measurement of low mass single diffractive dissociation, $p+p\rightarrow p+p^* \rightarrow p + X$,
where $X$ is a system of particles with typically $M(X) \sim$ few GeV. This physics is not possible with the
central detectors, as the hadrons coming from the fragmentation of $X$ typically have forward (longitudinal) 
momenta $\sim$ TeV/c and
transverse momenta $p_T \lsim$ 1 GeV/c. The FSC can not reconstruct the forward primary hadrons, but the patterns of their 
signals can be compared with simulations of soft diffraction to test the models. Such data will
strongly constrain existing models of diffractive processes and the $\mathrm{eff}$ects of rescattering and spectator parton interactions, see
for instance Ref.~\cite{kmr2}. This information is needed in order to understand the rapidity gap dynamics and gap survival
probability. No special running is needed; the standard low-$\beta$ running can be used, but 
most $\mathrm{eff}$ectively with low luminosity per
bunch crossing such that $\langle n \rangle \lsim$ 3. With 25 ns bunch spacing (2808 bunches), for $\sigma_{inel}$ = 80 mb, this
means luminosity $L < 4 \times 10^{32}$ cm$^{-2}$s$^{-1}$. Even when a store starts at $L= 10^{33}$ cm$^{-2}$s$^{-1}$,
after a few hours this condition will probably be met. 

The purity of the diffractive data sample (and hence the statistics for a given trigger
bandwidth) will be higher with one FSC arm in veto. The Collider Detector at Fermilab, CDF, 
installed a similar set of counters (beam
shower counters, BSC) used in veto at level-1 for some physics (central exclusive production)~\cite{cdfex}. 
These were pairs of scintillation counters, closely surrounding the beam pipe, at each of three (four on one side) locations,
6.6m, 23.2m, 31.6m (and 56.4m) from the intersection point. The closest counters had acceptance for primary particles with $5.4
< |\eta| < 5.9$, and were preceded by two radiation lengths of lead to convert photons. The other counters were behind
quadrupoles, electrostatic separators and (for the last counter) a dipole magnet. These only detected showers produced by
particles in the beam pipe and surrounding material. Together they covered 5.4 $< |\eta| <$ 7.4. 
The Tevatron beam has $y(p)$ = 7.65, where rapidity $y(p)$ = ln $\frac{\sqrt{s}}{m(p)}$. They were used $\mathrm{eff}$ectively both in level-1 triggers, and to tag events with proton dissociation. 
CDF also had a set of Roman pot detectors, with tracking,
to measure diffractively scattered antiprotons. It was found that even when $\langle n \rangle <$ 1, high mass single diffraction
studies ($X$ = dijets, $W,Z$ + hadrons) with a $\bar{p}$ track are dominated by pile-up (the $\bar{p}$ and $X$ being from different collisions)
unless a rapidity gap in the $\bar{p}$ direction is required to veto pile-up. Note however that in central exclusive production with
\emph{both} protons
detected ($p+p \rightarrow p+X+p$) 4-momentum conservation and precision (relative) timing of the protons enable physics to be done even
with $\langle n \rangle \gsim$ 20~\cite{mgafp420}.

In the following, a physics motivation is given for single diffractive excitation studies, and especially for central exclusive
production. We then present $\mathrm{eff}$iciency calculations.

\section{Single Diffractive Excitation}

Single diffractive excitation, SDE, is the process $p+p \rightarrow p + X$, where $``+"$ represents a large ($\gsim$ 3 units) rapidity gap,
meaning \emph{no hadrons} in pseudorapidity $\eta = -$ln tan$\frac{\theta}{2}$ between the outgoing proton and the diffractive 
system $X$. (Strictly, true rapidity $y = \frac{1}{2}$ln$\frac{E + p_z}{E-p_z}$ should be used, but for practical reasons $\eta$ is
usually considered to be an acceptable approximation.) 

The FSCs cover a crucial rapidity region between the zero degree calorimeters~\cite{zdc}, ZDC, in CMS and ATLAS (which only detect neutrons and photons
produced close to $\theta = 0^\circ$), CASTOR~\cite{castor}, and the TOTEM detectors T2~\cite{totem}. The dependence of $M(X)$ on
rapidity gap size $\Delta\eta$ is $\left( \frac{M(X)}{\sqrt{s}}\right)^2 \sim e^{-\Delta\eta}$, which can be used to estimate the mass spectrum, after correcting for the
more detailed relationship using Monte Carlo expectations. Note that the true rapidity of the diffractively scattered proton is $y \sim
$ln$ \frac{\sqrt{s}}{m(p)}$ = 9.3 (9.6) at $\sqrt{s}$ = 10 (14) TeV, but in the forward region $\eta$ and $y$ 
can be very different. A diffractively scattered proton with $p_T$ = 0.25 (0.5) GeV/c has $\eta(p)$ = 10.9 (10.2) at
$\sqrt{s}$ = 14 TeV (and 3\%  less at $\sqrt{s}$ = 10 TeV). A particle with $p_T = 0, \theta = 0$ has $\eta = \infty$. 

\begin{figure}[t]
  \vspace{9.0cm}
  \includegraphics{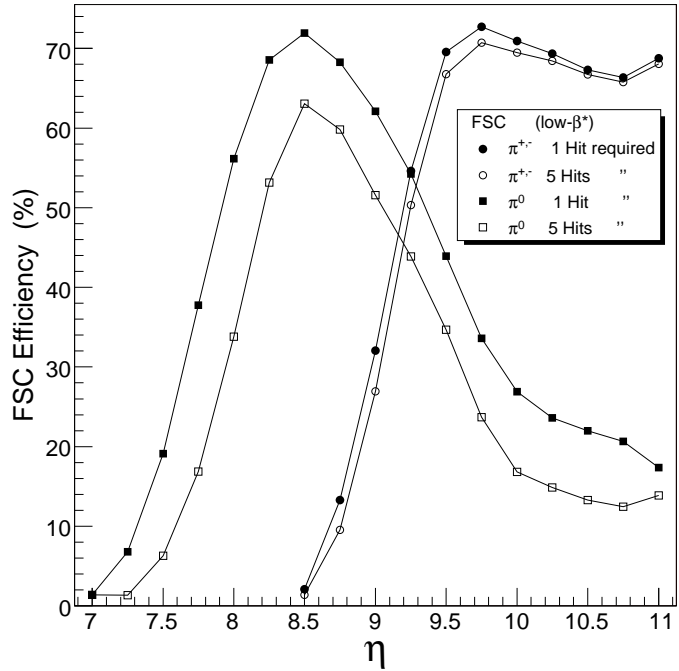}
  \caption{The ef$\mathrm{f}$iciency (\%) of the forward shower counters (FSC) for registering particle showers induced
  by primary $\pi^\pm$ and $\pi^0$ as a function of their pseudorapidity $\eta$ (low $\beta^*$ conditions).}
%\label{fig:cdfpu}
\end{figure}

%Figure~\ref{fig:cdfpu} demonstrates this with CDF (Tevatron) data, triggering on a forward antiproton.

%\begin{table}[htbp]
%\begin{center}
%\begin{tabular}{ccr}
%\hline
%\hline
%Location & z(mm) & $\Delta t$ (ns)\\
%1 & 59426 & 3.8 \\
%2 & 63751 & 0.0 \\
%3 & 68026 & 3.5\\
%4 & 72301 & 7.0 \\
%5 & 76576 & 10.5\\
%6 & 80851 & 11.0\\
%7 & 85126 & 7.5 \\
%\hline
%\hline
%\end{tabular}
%\end{center}
%\caption{Locations of counters on each (+ and -) beam pipe, and the time difference between incoming and outgoing bunch passes.}
%\label{ref:fscnumbers}
%\end{table}

\section{Central Exclusive Production (CEP)} 

In the following the reaction $p+p \rightarrow p+X+p$, where $X$ is fully measured and no other particles are produced, is
studied. We call this central exclusive production, CEP, or central exclusive diffraction, CED. 
The FSCs are used in the de$\mathrm{f}$inition of rapidity gaps (denoted $+$), where a gap means \emph{no hadrons}. In this
study, with no pile-up, proton tagging is not required. This is quite different from the FP420
proposal~\cite{mgafp420}, which is based on
a precision measurement of both forward protons. This reaction allows for a rich physics program~\cite{kmr2}, and in particular it provides a unique possibility
to study the two-gluon mediated color singlet interaction (i.e. using the LHC as a ``gluon-gluon collider'' $gg
\rightarrow gg$). In central exclusive production (or diffraction) the $gg\rightarrow q\bar{q}$ background is, in general, suppressed compared to $gg
\rightarrow gg$ due to color and spin factors. The selection rule $J_z=0$~\cite{kjz0}, valid for the CEP processes, further suppresses this
background by a factor $\propto \left( \frac{m(q)}{E_T}\right)^2$, where $m(q)$ is the quark mass and $E_T$ is the transverse energy of the jet. By
excluding inclusive central diffraction with relatively large proton transverse momenta ($p_T \gsim 0.7$ GeV/c), effective ``gluon
collider'' conditions are reached~\cite{kmr2}. As a result, detailed studies of pure high energy gluon jets can be made.
To guarantee the purity, it is essential to reject central \emph{inclusive} diffraction (CID) background, where one or both protons
dissociates into a multiparticle state, or the jets are accompanied by unassociated particles. 
These are ``direct backgrounds'', not a pile-up effect. The FSC would effectively veto the dissociation
background, as most of the fragmentation products hit the beam pipes and make showers, which are detected. 

The physics program includes a search for the production of mesonic states, such as glueballs, hybrids, heavy
quarkonia $\chi_c,\chi_b$ in double pomeron, $I\!\!P I\!\!P$, reactions as well as photoproduction: $\gamma I\!\!P \rightarrow J/\psi,
\psi(2S), \Upsilon$ ($I\!\!P$ denotes the pomeron). A strong coupling for the reaction $gg \rightarrow M$ for mesons
$M$ is expected as a result of two gluon exchange. A rapidity gap trigger on both arms could be used for this
study. The quantum numbers of the central state are restricted: the process is a ``quantum number $\mathrm{f}$ilter". The $t$-channel
exchanges over the large ($\Delta y \gsim$ 5 units) rapidity gaps can only be color singlets with charge $Q=0$ and
spin $J$ (or effective spin $\alpha(t)$) $\geq$ 1. Known exchanges are the photon $\gamma$ and the pomeron $I\!\!P$.
The gluon satis$\mathrm{f}$ies the $J = 1, \ Q = 0$ requirements, but it is not a color singlet; however one or more additional gluons
can cancel its color and form a pomeron. Another possible, but not yet observed, exchange in QCD is the odderon, $O$,
a negative C-parity partner to the $I\!\!P$ consisting of at least three gluons. This physics program includes
sensitivity to odderon exchange. States with $J^{PC} = 1^{--}$ such as the 
vector mesons ($V$) $J/\psi$ and $\Upsilon$ are
produced primarily by photoproduction, $\gamma I\!\!P \rightarrow V$. They can also be produced by odderon exchange,
$O I\!\!P \rightarrow V$, the signature for which would be a higher cross section for exclusive vector meson
production than that expected from photoproduction (and as predicted from HERA data, 
where the $O$ is absent), and with vector mesons having higher $p_T$
(on average). This is because on average $p_T(O) \gsim p_T(I\!\!P) \gg p_T(\gamma)$.

There is a long list of hadronic states which can be studied, pro$\mathrm{f}$iting from the quantum number $\mathrm{f}$ilter.
Glueball, $G$, spectroscopy, perhaps especially in the $G \rightarrow \phi\phi$ channel, and the exclusive production
of hyperon pairs, e.g. $I\!\!PI\!\!P \rightarrow \Lambda\bar{\Lambda}, \Sigma\bar{\Sigma}$ are examples out of many
that have not yet been studied by this technique. Many studies were done at the SPS ($\mathrm{f}$ixed target), but the energy was
too low for $I\!\!PI\!\!P$ dominance. (One can not have two rapidity gaps $\Delta y \gsim$ 3 units, with (say) 1
central unit for particle production, unless $\sqrt{s} > M(p) \times e^{3.5} = 31$ GeV, i.e. $p_{beam}$ = 546 GeV/c.)
The highest $\sqrt{s}$ at which the exclusive hadronic processes such as $I\!\!PI\!\!P \rightarrow \pi^+\pi^-, K^+K^-,
p\bar{p}$ and $\pi^+\pi^-\pi^+\pi^-$ were studied was $\sqrt{s}$ = 63 GeV~\cite{akes} at the CERN ISR. (Actually
$\alpha\alpha \rightarrow \alpha + \pi^+\pi^- + \alpha$ was measured at $\sqrt{s}$ = 126 GeV, but the
$\alpha$-particles had the
same rapidity as the $p$. Indeed the $M(\pi^+\pi^-)$ spectrum was the same as in $pp$ collisions at $\sqrt{s}$ = 63 GeV.) The
$\pi^+\pi^-$ mass spectrum showed striking structures: a probable $\sigma(600)$ broad enhancement, a narrow $f_0(980)$
and a clear dip around 1.6 GeV/c, which has still not been fully understood but may be a manifestation of a glueball state. 

A valuable reaction for calibrating forward spectrometers, e.g. as in FP420~\cite{mgafp420}, is 
the production of exclusive
di-leptons from $\gamma\gamma$ interactions and from $\Upsilon$ decay. A level-1 trigger, based on two muons or two
electromagnetic (EM) clusters and vetoing on the FSC counters, ZDC, and (in CMS) T1 and T2 detectors and the HF 
calorimeters, will select interactions with very large rapidity gaps and no pile-up. The rate of such events will be
acceptable even with a low ($\sim$ 4 GeV) $E_T$ cluster threshold and a similar low-$p_T$ threshold for muons, thus including
$\Upsilon \rightarrow e^+e^-, \mu^+\mu^-$ (for low $p_T$ $\Upsilon$). The CDF observations of exclusive lepton pairs and
charmonium states~\cite{cdfex} have been made possible thanks to their beam shower counters (BSCs).

\begin{figure}[t]
  \vspace{9.0cm}
  \includegraphics{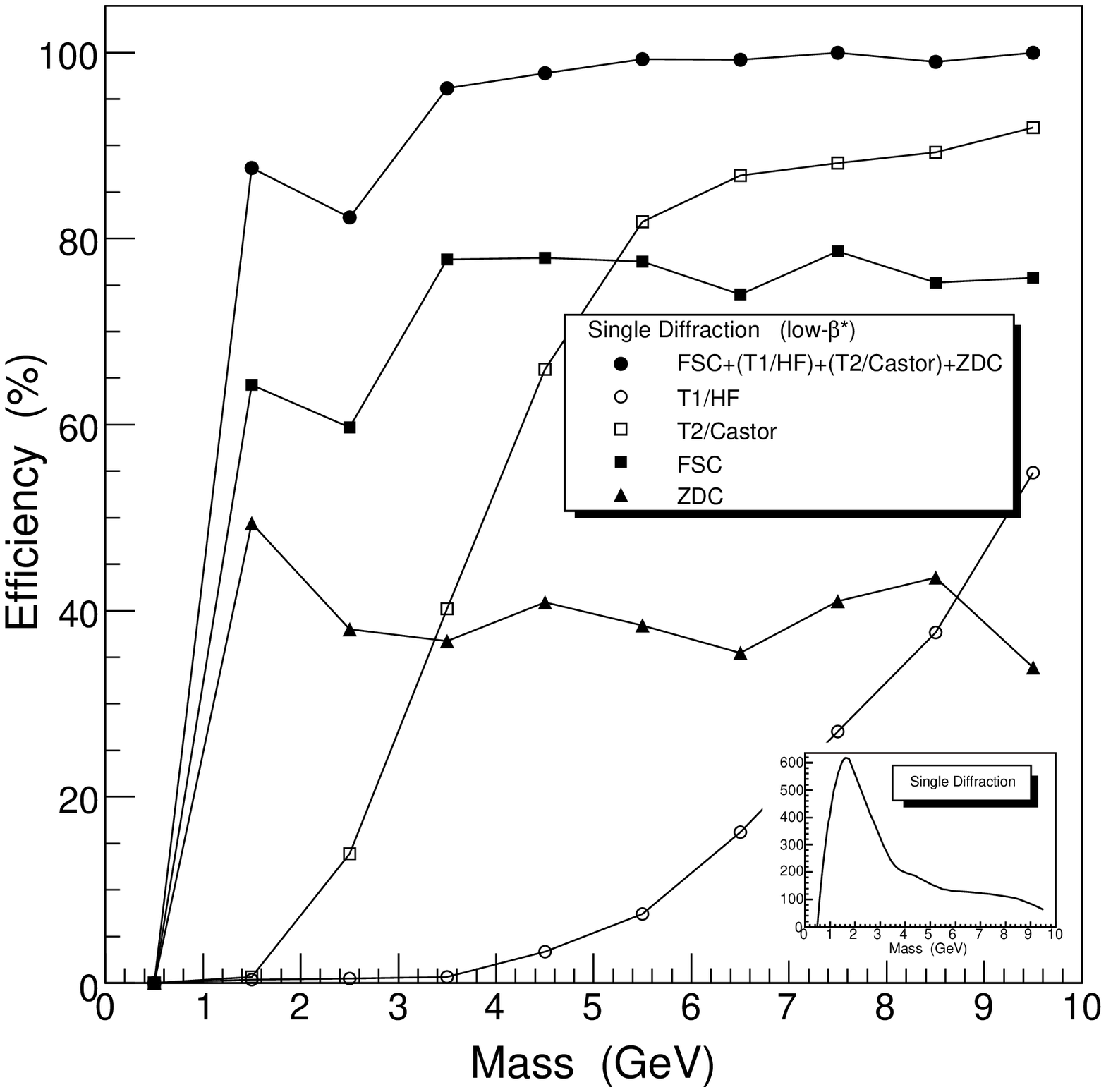}
  \caption{The detection ef$\mathrm{f}$iciencies for single diffrative events simulated by \textsc{pythia6.2} as a
  function of the diffrative mass. We required at least $\mathrm{f}$ive charged particles (``hits") in any of the forward shower 
  counters, or at least one track in the $\eta$ region covered by T1/HF or T2/CASTOR, or a minimum
  energy deposit in the ZDC (see text). }
%\label{fig:cdfpu}
\end{figure}

More generally, properties of central systems that have not previously been measured will be examined. In addition to low
mass exclusive central states (with all particles reconstructed, and mass $M(X) \lsim$ 10 GeV/c$^2$), the same trigger will
collect high mass double pomeron events, with $M(X)$ up to about 100 GeV/c$^2$. The jet content of such events, 
and in particular the subset of exclusive di-jets, is a valuable probe of the
parton constituents in the pomeron. Most such jets should be gluon
jets. One can compare $I\!\!P+I\!\!P$
collisions at $M(X)$ with $pp$ collisions at $\sqrt{s} = M(X)$. Some differences may be: a larger content of $\eta$ and
$\eta'$ mesons and a smaller baryon fraction because of the higher glue content. Pomerons should have a smaller
transverse size than
protons, and this may manifest itself in relatively more double parton scattering ($2 \times (gg \rightarrow JJ))$ and
perhaps also in Bose-Einstein correlations, which measure the size of the pion emission region. Double parton scattering, seen in
4-jet events having two pairwise-balancing dijets, is a probe of the (unintegrated) two-gluon density $G_2(x_i, x_j)$. Other ideas for early
LHC running using forward detectors for rapidity gap based physics have been presented~\cite{kmr2}. Furthermore the proposed
FSCs would provide information of value to the FP420 initiative~\cite{mgafp420} to place leading proton detectors at $\pm$240 m and $\pm$420
m from intersection Points 1 (ATLAS) and 5 (CMS). 

\section{Ef$\mathbf{f}$iciency for detecting rapidity gaps and for rejecting background}

 The FSC locations are in front of and behind the separation dipole D1,
and also between each of the six MBXW elements of D1 (see Figure 1). In addition counters can be placed 15 m, 35 m, and
55 m beyond D1, making a total of 10 detectors on each side. 
We have included in our ef$\mathrm{f}$iciency calulations the T1, T2, HF, CASTOR (one side only) and ZDC detectors. The TOTEM tracker
T1 and the forward calorimeter HF span the region $3 < |\eta| < 5$. Tracker T2 and the CASTOR calorimeter cover $5 < |\eta|
< 7$. The Zero Degree Calorimeter, ZDC, is in-between the two beam pipes just beyond their separation, and detects only
neutral particles (mainly $\gamma$ and neutrons) with $|\eta| > 8.5$. The program \textsc{geant}~\cite{geant} has been used
to simulate the beam line, including the beam pipes, beam screens, and magnetic elements. The running condition is for the
standard low-$\beta$ con$\mathrm{f}$iguration, $\beta^* = 0.55$ m; no special running is required for this program. 
However its
ef$\mathrm{f}$iciency will become less than 5\% when the average number of 
inelastic collisions per bunch crossing, $\langle n
\rangle > 3$. 
 
 \begin{figure}[t]
  \vspace{9.0cm}
  \includegraphics{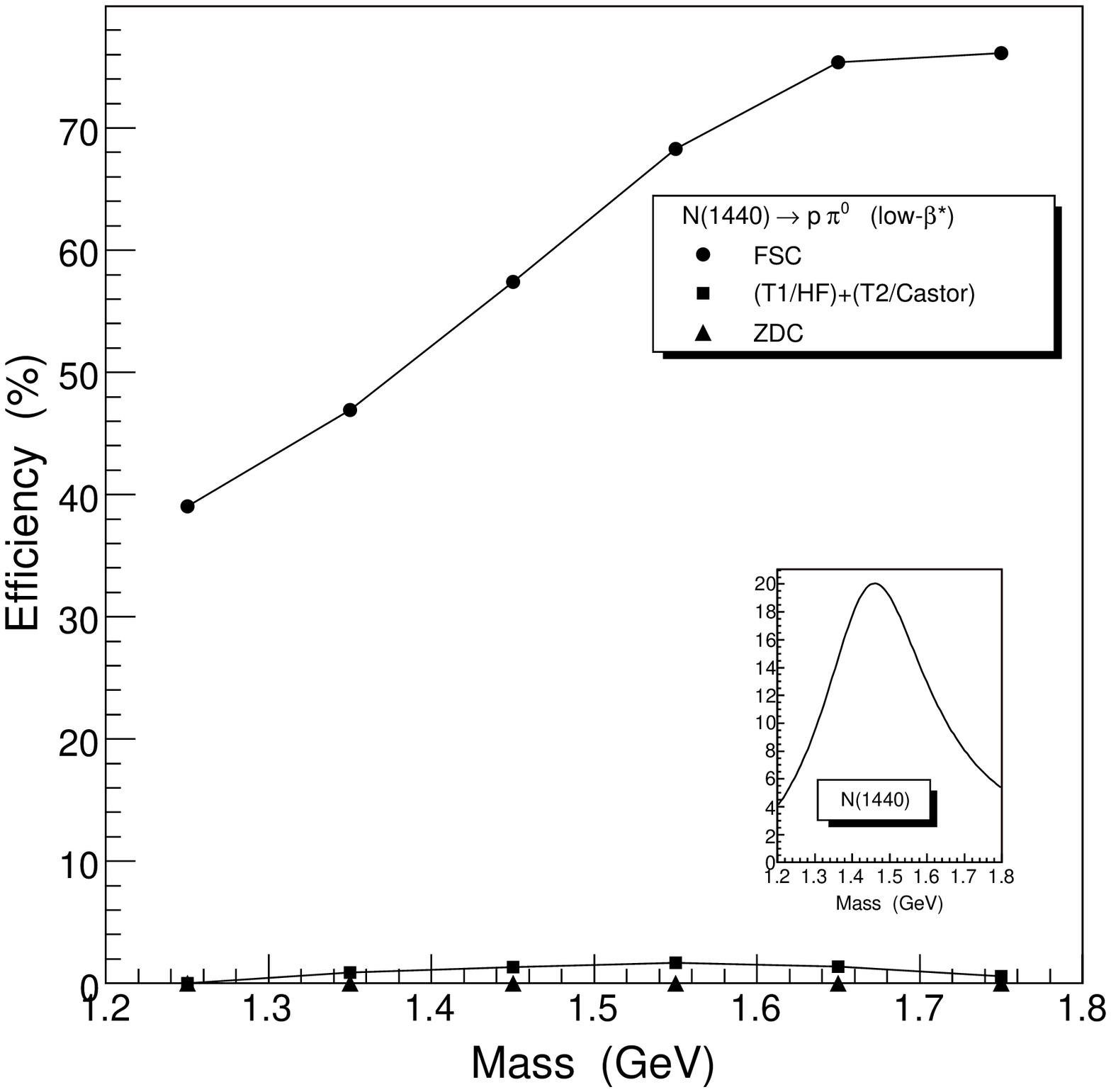}
  \caption{The detection ef$\mathrm{f}$iciency for single diffractive events with $N^* \rightarrow p+\pi^0$ as a
  function of diffractive mass. We required at least $\mathrm{f}$ive hits in any of the forward shower 
  counters, or at least one track in the $\eta$ region covered by T1/HF or T2/CASTOR, or a minimum
  energy deposit in the ZDC (see text).}
%\label{fig:cdfpu}
\end{figure} 

 \begin{figure}[t]
  \vspace{9.0cm}
  \includegraphics{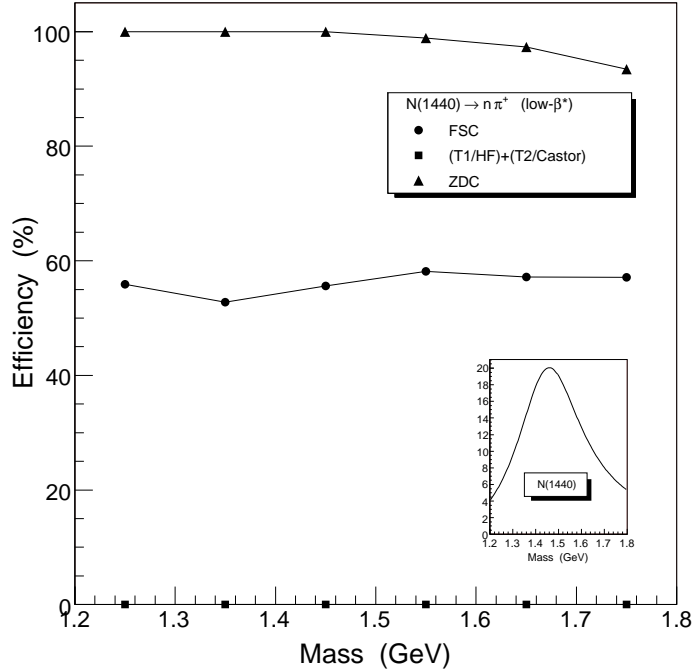}
  \caption{As Fig.5 for $N^* \rightarrow n+\pi^+$.}
%\label{fig:cdfpu}
\end{figure} 

\section{Single particle ef$\mathbf{f}$iciency of FSCs}

The FSC detection ef$\mathrm{f}$iciency for incident particles ($\pi^\pm,\pi^0$) was calculated as a function of
pseudorapidity $\eta$. The requirement was at least one particle (``hit") (alternatively at least $\mathrm{f}$ive) in any of the FSC
counters. A transverse momentum ($p_T$) distribution of the form $e^{-6.7 p_T^2}.dp_T^2$ was assumed for the
incident primary particles, corresponding to that obtained from \textsc{pythia 6.2}~\cite{pythia}. The ef$\mathrm{f}$iciency of the FSCs for
detecting charged particles from showers induced by the primary $\pi^\pm$ and $\pi^0$ is shown in Figure 3. For
charged pions the e$\mathrm{ff}$iciency is $\sim$70\% for $|\eta| >$ 9.5, and it is nearly independent of the number of hits,
at least for 1 - 5 hits per detector plane. For $\pi^0$ between 8 $< |\eta| <$ 9.3 it exceeds 65\% (50\%) when at
least 1 (5) hits are required. From the results presented in the following sections, this is suf$\mathrm{f}$icient for the
anticipated physics studies.

\section{Single diffraction detection e$\mathbf{ff}$iciency}

The detection ef$\mathrm{f}$iciencies for single diffractive excitation, as simulated with \textsc{pythia 6.2}, were
calculated as a function of the diffractive mass. They were also calculated with \textsc{phojet 6.2}~\cite{engel}
and found to agree with those from \textsc{pythia}, within statistics.
We required at least $\mathrm{f}$ive hits in any FSC counter, or a track or signal in the $|\eta|$ region covered by T1, T2, HF, CASTOR
or the ZDC. A ``signal'' in the HF or CASTOR is de$\mathrm{f}$ined as an energy deposit above 15 GeV, or above 500 GeV in the
ZDC. The 500 GeV is nominal. Once data are obtained at low luminosity with a zero-bias (bunch crossing) trigger, it will be possible to
optimise the cuts, for each detector, to provide the best separation between events with a true gap (no
particles) and with particles. As in the CDF analysis, one can divide the zero-bias events into two classes: those
apparently empty (no tracks and no large electromagnetic clusters) and those with interactions. For such studies it
is necessary to have zero-bias data recorded, especially at low luminosity when the fraction of empty
crossings is not too small. The e$\mathrm{ff}$iciencies as a function of diffractive mass for these conditions, along with what would be obtained using
only T1/HF, T2/CASTOR, FSC or ZDC detectors are shown in Figure 4. The ef$\mathrm{f}$iciency is $>$90\% for the lower mass  region,
and approximately 100\% for masses above 10 GeV (not shown). Approximately 25\% of the single diffractive cross section
is for masses below 10 GeV (at $\sqrt{s}$ = 14 TeV). Forward multiparticle states from central inclusive diffraction
reactions would have similar detection e$\mathrm{f}$iciency.

 \begin{figure}[b]
  \vspace{9.0cm}
  \includegraphics{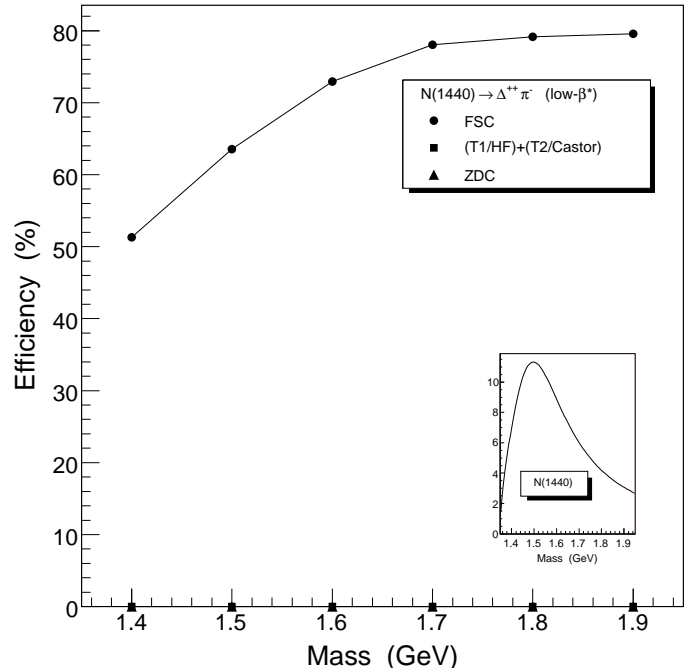}
  \caption{As Fig.5 for $N^* \rightarrow \Delta^{++} + \pi^-$.}
%\label{fig:cdfpu}
\end{figure} 

Simulations have also been made for exclusive diffractive baryon resonance production, such as $p+p \rightarrow p +
N^*(1440)$ with $N^* \rightarrow p+\pi^0, n+\pi^+$,  or $\Delta^{++} + \pi^-$. 
In Figures 5-7 the ef$\mathrm{f}$iciencies for
detecting these $\mathrm{f}$inal states are shown as functions of the diffractive mass. For $N^* \rightarrow p+\pi^0$ the average
e$\mathrm{f}$iciency is 70\% (Figure 5), for $N^* \rightarrow n+\pi^+$ it is close to 100\% (Figure 6), and for $N^*\rightarrow
\Delta^{++}+\pi^-$ it is about 70\% (Figure 7).

An approximate calculation of the diffractive mass can be made through its relation to the size of the rapidity gap adjacent to the
scattered proton, although this has some model dependence. The relation depends on the $p_T$ distribution (and hence $\langle
p_T \rangle$) of the produced particles. The ``adjacent rapidity gap'' is de$\mathrm{f}$ined as the gap between the diffractive
proton (close to the beam rapidity, $y_{beam} = 9.6$ at $\sqrt{s}$ = 14 TeV) and the nearest particle in rapidity.
Larger rapidity gaps correspond to smaller diffractive masses. The approximate correspondence between the diffractive
mass $M$ and the (pseudo)rapidity gap $\Delta\eta$ is $\left( \frac{M(X)}{\sqrt{s}}\right)^2 \sim e^{-\Delta\eta}$. From the theoretical point of view it is
more instructive to consider the distribution $\frac{d\sigma_{SD}}{d\eta'}$, where $\eta'$ is the position of the edge of
the gap. Although asymptotically $\frac{dM^2}{M^2}= d\eta'$ due to angular ordering of gluon emission, different
contributions to $\sigma_{SD}$ are separated from each other in $\eta'$ rather than in $M$. 

 \begin{figure}[t]
  \vspace{9.0cm}
  \includegraphics{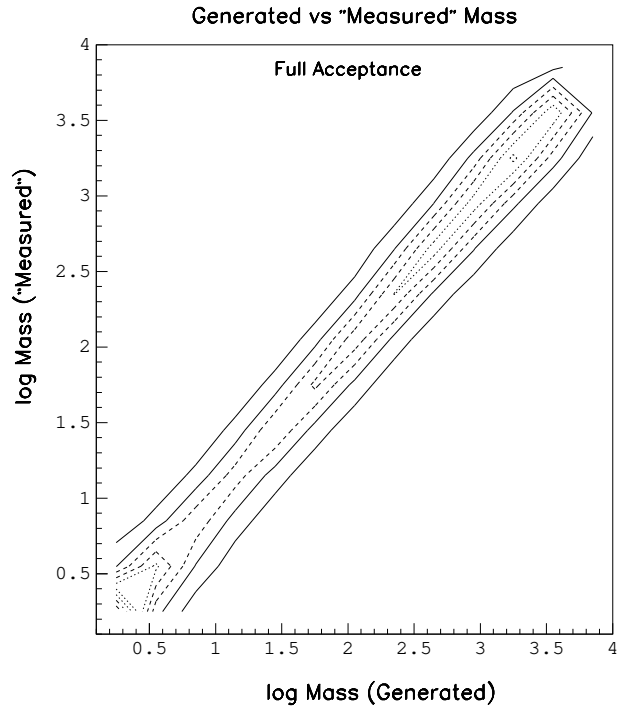}
  \caption{The diffractive mass reconstructed (``measured'') from the width of the rapidity gaps vs. the
  true (generated) mass. The lines show contours of equal density. All forward detectors including the FSC are assumed.}
%\label{fig:cdfpu}
\end{figure} 

To provide a more precise (although model dependent) measurement, the \textsc{pythia} program has been used to determine
the correlation between the diffractive mass and the size of the rapidity gap. Figure 8 shows the true diffractive mass $M$
versus $\Delta\eta$ as determined by this method. To account for the measurement resolution, a Gausssian 
spread with $\sigma=$ 10\%
has been added to the actual rapidity value. This is more than one unit at the largest values considered, and is
considered to be an overestimate. Figure 9 shows the actual (generated) diffractive mass together with that calculated by the
above method, for two cases: (a) for full $\eta$ coverage, and (b) for the limited $\eta$ range $|\eta| <$ 4.7, i.e. the
nominal CMS coverage. Clearly the wider the range of rapidity covered, the more accurately the diffractive mass can be
determined from the rapidity gap size $\Delta\eta$.

Determination of the diffractive mass on an event-by-event basis from the dependence on $\Delta\eta$ is imprecise for
low masses, $M \lsim$ 5 GeV/c$^2$. For single diffraction one relies largely on the FSC in this mass range. For central
exclusive production the central detectors measure the mass with relatively good precision. The contribution to the
total single diffractive cross section in this mass range from the FSC data only is shown by the $\mathrm{f}$irst solid point and
horizontal dashed line in Figure 9, representing the average measurement for the $\mathrm{f}$irst two bins. Further analysis of the
FSC data should improve these results.

\begin{figure}[t]
  \vspace{9.0cm}
  \includegraphics{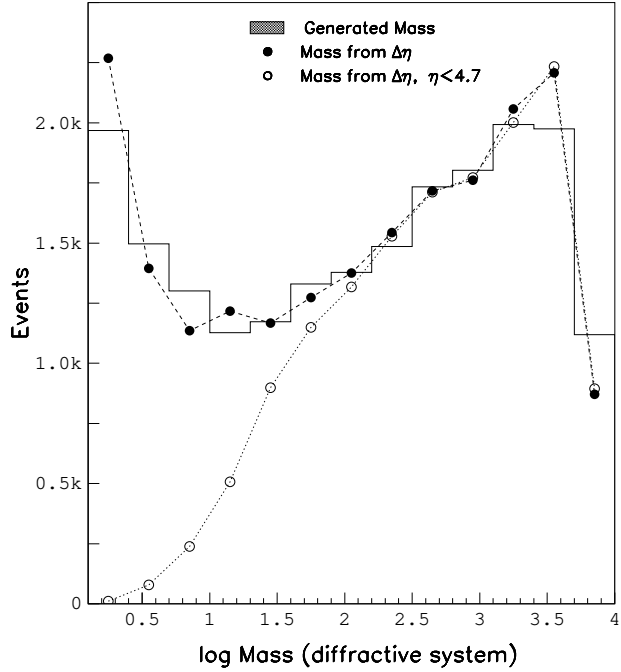}
  \caption{Distribution of the actual (generated) diffractive mass, log $M(X)$, together with that
  calculated using the rapidity gap measurement for two cases: (a) full $\eta$ coverage, and (b)
  for a limited $\eta$ range, $|\eta| < 4.7$. The $\mathrm{f}$irst solid point with the horizontal dashed line
  indicates the contribution from the FSC data.}
%\label{fig:cdfpu}
\end{figure} 

The e$\mathrm{ff}$iciency of the FSC for detecting forward diffractive systems is high. However it is not 100\%, and as a result the
SDE and CEP studies will contain some background. A subtraction technique can be used to estimate this background and
remove it. Data can be taken (a) with, and (b) without the use of the FSC for rapidity gap detection, with T1/HF and
T2/CASTOR in veto in both cases. Case (b) includes increased background and characterises the FSC ine$\mathrm{f}$iciency. One can
also (off-line) measure the content of individual FSC counters, which cover different $\eta$-ranges; this provides more
differential tests of the diffractive event simulation. Measuring the various rates, with knowledge of the FSC
ef$\mathrm{f}$iciencies, the background contributions can be estimated and subtracted for different situations (e.g. different
$M(X)$). Correlations between the counters can be determined and compared with expectations. A valuable check will be the
independence of all the measured cross sections on the instantaneous luminosity.  

\section{Central exclusive production detection e$\mathbf{ff}$iciency}

Central exclusive production has two leading protons (not detected without Roman pots or similar devices) adjacent to
rapidity gaps of $\gsim$ 4 units. The $t$-channel 4-momentum exchanges can be carried by photons, pomerons or (not yet observed)
odderons. Therefore the central state $X$ can result from $\gamma\gamma \rightarrow X, \gamma I\!\!P \rightarrow X$ or
$I\!\!PI\!\!P \rightarrow X$. As the electromagnetic coupling is much smaller than the strong coupling, the $\gamma\gamma
\rightarrow q\bar{q}$ process is an insigni$\mathrm{f}$icant background to $I\!\!PI\!\!P \rightarrow$ hadrons, 
but $\gamma\gamma$ collisions are cleanly observed in
exclusive $X = e^+e^-,\mu^+\mu^-$~\cite{cdfex}. (At high luminosity $\gamma\gamma \rightarrow W^+W^-$ and possible
$\tilde{l}^+\tilde{l}^-$ may be observed, in the presence of pile-up and with forward proton measurement.) For low mass exclusive states, e.g.
$X = \phi\phi$, note that $p_T(X)$ is lower on average from $\gamma\gamma$ exchanges than from $I\!\!PI\!\!P$.

Central exclusive production (CEP) was simulated using \textsc{phojet1.1}~\cite{engel} to generate the central diffractive
mass, and \textsc{pythia} to decay the central system into a gluon-gluon dijet. The detection e$\mathrm{f}$iciencies for central
diffractive events were calculated as functions of the central mass $M(X)$. We required less than $\mathrm{f}$ive hits in any FSC
counter and no tracks in the $\eta$ regions covered by the T1/HF and T2/CASTOR detectors. 

For central \emph{inclusive} diffraction events, we studied the probability of having at least $\mathrm{f}$ive 
hits in the FSC (in both arms),
and the probability of having at least one track in the T1/HF or T2/CASTOR regions, as a function of the central
mass $M(X)$. Requiring an FSC veto is seen to be ef$\mathrm{f}$icient, and requiring a T2/CASTOR veto is ef$\mathrm{f}$icient for central masses
$M(X) \gsim$ 120 GeV/c$^2$, but requiring a T1/HF veto would reject some 25\% to 35\% of these events, creating a bias.
However if one is interested only in the subset of central diffractive production with no particles beyond $|\eta|$ = 3, the
T1/HF veto would be included. 

We have also made simulations of the reactions $p+p \rightarrow p + X + p^*$ and $p+p \rightarrow p^* + X + p^*$, where
$p^*$ is a forward diffractive system and $X \rightarrow gg$. These reactions are similar to the ``quasi-elastic'' case
where the protons do not dissociate, and the study shows similar results.

As shown in Ref.~\cite{kmr2}, the cross section of central diffractive production $\frac{d\sigma_{CD}}{d\eta_1d\eta_2}$ 
is particularly sensitive to the models of soft diffraction, and these measurements will provide valuable information on
the parton content and sizes of various diffractive eigenstates. Measurement of the rapidity gap survival probability,
$\hat{S}^2$, which determines the diffractive cross sections, is important for understanding strong interaction
processes~\cite{gribov,kmr2}. The present estimates are based on model calculations and $\hat{S}^2$ must be
experimentally measured.

\begin{figure}[t]
  \vspace{9.0cm}
  \includegraphics{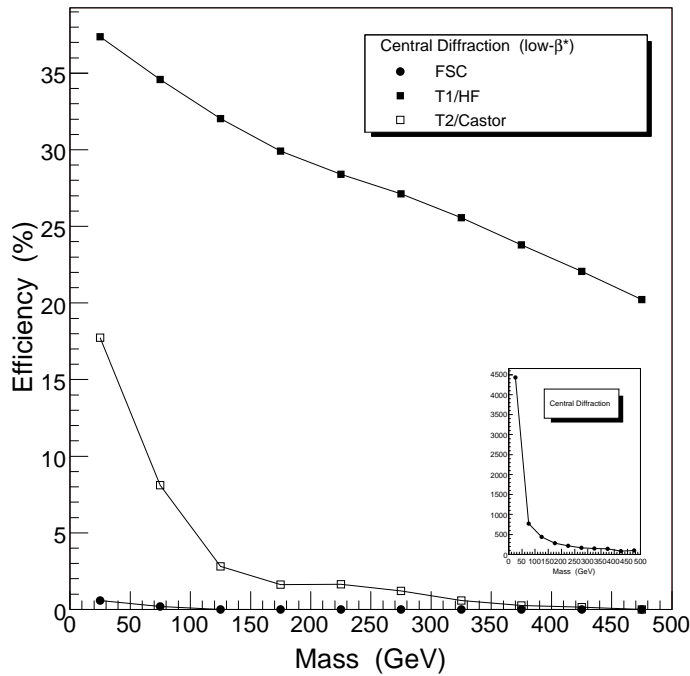}
  \caption{Veto efficiencies for central exclusive diffractive, CED, events simulated by \textsc{phojet1.1/pythia6.2} as
  a function of the central mass $M(X)$. Separately for the FSC, for T1/HF, and for T2/CASTOR, we call ``veto efficiency'' the
  probability that central events will have charged particles in both forward arms.}
%\label{fig:cdfpu}
\end{figure} 

\section{Non-diffractive detection ef$\mathbf{f}$iciency}

The detection ef$\mathrm{f}$iciency for non-diffractive events was calculated as a function of the charged multiplicity. Requiring at
least $\mathrm{f}$ive hits in any of the FSC counters, and/or at least one track in any region covered by T1/HF or T2/CASTOR, we
$\mathrm{f}$ind the e$\mathrm{ff}$iciency to be close to 100\%, except for very low multiplicity events. The FSC 
alone have an ef$\mathrm{f}$iciency for non-diffractive events of about 90\%.

\section{Conclusions}

Because of limited forward detector coverage, measurements of single diffractive and central diffractive cross sections in
hadron-hadron collisions are more limited at higher $\sqrt{s}$ values than that of the CERN ISR ($\sqrt{s} \leq$ 63 GeV).
The published SDE cross sections at the SPS and Tevatron~\cite{tev} were obtained by extrapolation from
data collected in limited $p_T$ and $\eta$ regions. At the LHC, diffractive cross sections can be measured with the
addition of forward shower counters, FSC, to the present CMS or ATLAS detectors to cover the lowest diffractive masses,
below $\sim$ 10 GeV/c$^2$. With the proposed detector arrangement, valuable new data can be obtained by tagging single and
central diffractive processes. The e$\mathrm{ff}$iciency calculations show that one can use the two-gluon mediated 
color-singlet
interaction as a ``gluon collider''. The ef$\mathrm{f}$iciency of the FSC system for detecting rapidity gaps is shown to be adequate for
the proposed studies of single- and central-diffraction.

The FSC could also serve as a luminosity monitor by measuring the East-West concidence rate (or equivalently the probability of bunch
crossings with no forward particles), as well as monitors of beam conditions. 

\section{Acknowledgements}

We thank D.Svododa and R.Hall-Wilton for information on the LHC beam line and discussions.

\end{document}